\documentclass[aps, prx, superscriptaddress, nofootinbib, reprint, twocolumn, floatfix, preprintnumbers]{revtex4-2}
\usepackage{latexsym, amsmath, amsfonts, amssymb}

\usepackage{graphicx}

\usepackage{comment}

\usepackage{subcaption}

\usepackage{hyperref}
\hypersetup{colorlinks = true, linkcolor = blue, anchorcolor = blue, citecolor = blue, filecolor = blue, urlcolor = blue}
\usepackage{bbm}
\usepackage{mathrsfs}
\usepackage{array} 
\usepackage{xcolor}
\usepackage{slashed}
\usepackage{bbold}
\usepackage[normalem]{ulem}
\renewcommand{\arraystretch}{2.5}

\usepackage{tikz}
\usepackage{tikz-3dplot}
\usetikzlibrary{decorations.pathmorphing} 
\usepackage{enumitem}

\setlist[itemize]{leftmargin=1.25em}
\setlist[description]{leftmargin=0em, itemsep=0em}
\setlist[enumerate]{leftmargin=1em, itemsep=0em}

\newcommand{\ds}{\displaystyle}

\newcommand{\be}{\begin{equation}} \newcommand{\ee}{\end{equation}}
\newcommand{\bea}{\begin{equation} \begin{aligned}} \newcommand{\eea}{\end{aligned} \end{equation}}

\newcommand{\cC}{\mathcal{C}}

\newcommand{\cF}{\mathcal{F}}

\newcommand{\cM}{\mathcal{M}}
\newcommand{\cN}{\mathcal{N}}
\newcommand{\cO}{\mathcal{O}}

\newcommand{\cR}{\mathcal{R}}
\newcommand{\cS}{\mathcal{S}}

\newcommand{\bZ}{\mathbb{Z}}

\newcommand{\simAdS}{\underset{z \to 0}{\sim}}
\newcommand{\simAdSzzero}{\underset{z_0 \to 0}{\sim}}
\newcommand{\widehatBox}{\widehat{\vphantom{A}\Box}}

\DeclareMathOperator{\Tr}{Tr}

\newcommand{\AdS}{\text{AdS}}

\definecolor{darkgreen}{rgb}{0.1,0.6,0.1}

\makeatletter
\renewcommand\@makecaption[2]{%
  \par
  \vskip\abovecaptionskip
  \begingroup
    \small\rmfamily
    \raggedright 
    \@make@capt@title{#1}{#2}\par
  \endgroup
  \vskip\belowcaptionskip
}
\makeatother

\begin{document}

\title{F-theorem for Quantum Field Theories in Anti-de Sitter Space \\
}

\newcommand{\OXFORD}{\affiliation{Mathematical Institute, University
of Oxford, Woodstock Road, Oxford, OX2 6GG, United Kingdom}}

\newcommand{\CERN}{\affiliation{Department of Theoretical Physics, CERN, 1211 Meyrin, Switzerland}}

\newcommand{\UNITS}{\affiliation{Dipartimento di Fisica, Universit\`a di Trieste, Strada Costiera 11, I-34151 Trieste, Italy
}}

\newcommand{\BOSTON}{\affiliation{Department of Physics, Northeastern University, Boston, MA, 02115, USA
}}

\newcommand{\INFN}{\affiliation{INFN, Sezione di Trieste, Via Valerio 2, I-34127 Trieste, Italy}}

\newcommand{\BEIJING}{\affiliation{Yau Mathematical Sciences Center, Tsinghua University, Jingzhai, Haidian District, Beijing, 100084, China
}}

\author{Davide Bason}
\BEIJING

\author{Christian Copetti}
\OXFORD

\author{Lorenzo Di Pietro}
\UNITS 
\INFN

\author{Ziming Ji}
\BOSTON

\author{Shota Komatsu}
\CERN

\begin{abstract}
We introduce a regularized free energy $\cF_{\AdS}$ for massive quantum field theories (QFTs) on Anti-de Sitter space (AdS). We conjecture this quantity to be monotonic under 
the renormalization group (RG) flow induced by boundary perturbations,
generalizing the known boundary $F$-theorem to non-conformal setups.
We test this conjecture in several examples and provide a proof in two dimensions. We also discuss applications to long-range critical points, obtaining bounds on the sphere free energy of long- and short-range Ising models in three dimensions.
\end{abstract}

\maketitle
\section{Introduction}
Understanding the space of $d$-dimensional Conformal Field Theories (CFTs) and the renormalization group (RG) trajectories connecting them is a central problem in quantum field theory (QFT). A key tool in constraining such flows is  monotonicity theorems, which assign an RG-decreasing quantity
 $\cF$ \cite{Zamolodchikov:1986gt,Cardy:1988cwa,Komargodski:2011vj,Jafferis:2011zi} to each CFT, thereby limiting the allowed RG trajectories.

Richer structures appear in the presence of boundaries or defects that preserve an $SO(p+1,1) \times SO(d-p)$ subgroup of the Euclidean conformal group. In this context, several boundary/defect monotonicity theorems have been proposed or proven, assuming that the bulk theory remains conformal \cite{Affleck:1991tk,Friedan:2003yc,Gaiotto:2014gha,Jensen:2015swa,Casini:2016fgb, Casini:2018nym, Wang:2020xkc,Wang:2021mdq,Cuomo:2021rkm, Casini:2023kyj,Kobayashi:2018lil}. This assumption is essential: counterexamples arise when both bulk and defect operators participate in the RG flow\footnote{However, see \cite{Arav:2024exg} for an interesting proposal in the context of supersymmetric monodromy defects.} \cite{Green:2007wr,Shachar:2024ubf}.

In this Letter, we extend this framework to massive, i.e. non-conformal, bulk QFTs\footnote{For monotonicity theorems for CFTs in AdS, see \cite{Giombi:2020rmc}. Our analysis differs in that we focus on \emph{non-conformal} bulk QFTs rather than CFTs.}. The key idea is to place the bulk QFT on a fixed Anti-de Sitter (AdS) background \cite{Callan:1989em}. Euclidean AdS$_{d+1}$ has isometry group 
$O(d+1,1)$, which acts as the conformal group on the AdS boundary; consequently, massive bulk dynamics can be probed  through conformal correlators of boundary operators \cite{Paulos:2016fap,Carmi:2018qzm,Komatsu:2020sag,Cordova:2022pbl,vanRees:2022zmr,vanRees:2023fcf,Meineri:2023mps,Levine:2023ywq,Lauria:2023uca,Antunes:2024hrt,Ghosh:2025sic}.
 Kinematically, this setup resembles the study of conformal boundary conditions in CFT (BCFT), despite the bulk being massive. To emphasize both the analogy and the distinction, below we refer to the boundary dynamics of a massive QFT in AdS as  {\it bCFT}.\footnote{We stress that with this name we are referring to a set of conformally covariant boundary correlation functions. Among the set of boundary operators there is no conserved stress tensor. In all other respects the bCFT obeys the usual CFT axioms such as discreteness of the spectrum, unitarity and crossing symmetry.}

There has been substantial recent interest in bCFTs, particularly for analyzing asymptotically free bulk theories. 
Varying the scale $\Lambda$ associated to bulk interactions, measured in units of the AdS radius $L$, enables a controlled interpolation between the weakly coupled UV regime $\Lambda L \ll 1$, and the strongly coupled flat-space IR regime $\Lambda L \gg 1$. From the bCFT perspective the dimensionless combination $\Lambda L$ labels a family of bCFTs. Building on earlier work \cite{Aharony:2012jf,Carmi:2018qzm},
this strategy has been applied to a variety of asymptotically free systems, including dynamical mass generation in large $N$ models \cite{Copetti:2023sya}, perturbative studies of onset of confinement \cite{Ciccone:2024guw,Ciccone:2025dqx}, analyses of confining flux tubes in Yang–Mills \cite{Gabai:2025hwf}, and the emergence of Seiberg–Witten geometry from weakly coupled $\cN=2$ gauge theory \cite{Bason:2025zpy}.  

Here, however, we focus on the dynamics at {\it fixed} 
$\Lambda L$. Even holding 
$\Lambda L$ constant, a nontrivial flow of boundary conditions may arise when the boundary interaction breaks AdS isometries or, equivalently, when boundary-localized deformations are introduced. From the bCFT viewpoint, these constitute RG flows between distinct bCFTs. A prototypical example is the “double-trace flow’’ \cite{Klebanov:1999tb,Witten:2001ua}, in which a free massive scalar 
$\phi$ flows between two AdS-invariant boundary conditions under a boundary $\phi^2$ deformation. 

The central claim of this letter is a conjecture that flows between bCFTs admit a monotonically decreasing function, given by the regularized free energy on AdS$_{d+1}$,
\be
\cF_{\AdS} = (-)^{s_d} \, \text{Re}\left( F_{\AdS_{d+1}}  - \kappa_d \, F_{S^{d+1}} \bigg\vert_{R = i L} \right) \, .\label{eq:Fads}
\ee
Here $F$ is the free energy $F= - \log Z $, $R$ is the radius of $S^{d+1}$, and $\kappa_d$ and $s_{d}$ are given by\footnote{We chose the sign $s_d$ so that the F-function is \emph{decreasing} for all dimensions, as in \cite{Giombi:2014xxa}, where a similar free energy was discussed for CFTs.}
\begin{align}
\begin{cases}
\text{odd }d: & \kappa_d =\frac{1}{2}\,,\qquad s_{d}=\frac{d+1}{2}\,,\\
\text{even }d: & \kappa_d =\frac{i}{\pi}\,,\qquad s_{d}=\frac{d}{2}\,.
\end{cases}
\end{align}
We refer to $\cF_{\AdS}$ as the {\it AdS free energy} and to its conjectured monotonicity  {\it AdS F-theorem}. The conjecture implies that the quantity $\cF_{\AdS}$ gives an organizing principle for the several possible bCFTs associated to a fixed bulk theory.

In \eqref{eq:Fads}, as in the BCFT analysis of \cite{Gaiotto:2014gha}, the bulk UV divergence is automatically subtracted by $F_{S^{d+1}}$. The bulk IR divergence instead has to be removed using techniques of holographic renormalization \cite{Bianchi:2001kw, Skenderis:2002wp}. 

We provide a proof of \eqref{eq:Fads} in $d=1$ following \cite{Friedan:2003yc,Cuomo:2021rkm} while providing evidence for higher $d$ in several examples. 
We also discuss a few applications including the derivation of bounds on the sphere free energy of short-range and long-range Ising fixed points \cite{Fisher:1972zz,Sak:1973oqx,Honkonen:1988fq,Paulos:2015jfa,Behan:2017emf, Behan:2017dwr, Benedetti:2024wgx,Behan:2025ydd} in two and three dimensions, which can be realized as massive scalars in AdS coupled to the boundary degrees of freedom. 

\section{The AdS free energy}
We start by tackling the problem of defining a scheme-independent, UV finite AdS free energy. The naive free energy $F_{\AdS} = - \log | Z_{\text{AdS}}|$ suffers from divergences of two types: IR divergences from the infinite AdS volume and UV divergences from short-distance bulk physics. IR divergences can be regulated by boundary counterterms, a procedure termed holographic renormalization \cite{Bianchi:2001kw, Skenderis:2002wp} in the context of AdS/CFT, which however can be applied more broadly to QFT in AdS. In this process, we add to the bare action IR divergent local counterterms on the cutoff surface $z = z_0$, with $z$ being the radial AdS coordinate (the AdS boundary lies at $z=0$). 
These counterterms are constructed from boundary data such as the boundary metric $h_{ab}$ and extrinsic curvature $K_{ab}$, as well as the restriction of bulk operators $\Phi$ to the cutoff surface.\footnote{In most studies stemming from the AdS/CFT correspondence, the holographic renormalization is performed around a semi-classical configuration. In this case, there is a well established methodology to derive all the needed counterterms \cite{Papadimitriou:2004ap}. See \cite{Banados:2022nhj} for extensions to higher orders in perturbation theory. Notice that the bulk UV divergences will affect the holographic counterterms, which will generally become UV-dependent.}  
 Power-law divergences $\sim z_0^{-n}$ can always be regularized in this manner. The remaining IR finite contribution, or alternatively the coefficient of the $\log(z_0)$ divergence, if present, are physically meaningful quantities. A typical example in which an IR divergent counterterm is needed is the case of a bCFT with a boundary relevant operator. In this case the counterterm is simply interpreted as the fine-tuning of boundary terms needed to define an AdS-symmetric boundary condition. 
 
 In addition, typically there are UV divergences, which depend on the bulk UV cut-off $\Lambda_{\rm UV}$. Crucially, these divergences are independent of the choice of boundary condition and can be regulated in two natural ways:
\begin{enumerate}
    \item We consider the difference between the free energies with two different boundary conditions $B_1$ and $B_2$:
    \be
     \cF_{\AdS}(B_1, B_2) = (-)^{s_d}(F_{\AdS}(B_1) - F_{\AdS}(B_2)) \, ,
    \ee
    at the same (fixed) AdS radius $L$. We have included a conventional sign $(-)^{s_d}$, as in \eqref{eq:Fads}, in such a way that the function will be \emph{decreasing} along boundary RGs. This can be interpreted in two equivalent ways: either as the statement that $ \cF_{\AdS}(B_1, B_2)$ is non-negative if there is a boundary RG starting from $B_1$ and ending in $B_2$, or by keeping $B_2$ as a fixed reference and looking at this quantity as a function of $B_1$. While bulk UV divergences cancel out in the difference, no canonical choice of reference boundary condition is available. In practice $\cF_{\AdS}(B_1, B_2)$ remains a convenient quantity to compute for checking the AdS F-theorem.
    \item Following a familiar BCFT procedure \cite{Gaiotto:2014gha}, we can also subtract bulk UV divergences on a suitably chosen closed compact manifold. For a BCFT this manifold is the sphere \cite{Giombi:2020rmc,Giombi:2025pxx}, and indeed a conformal theory in global AdS can be mapped to a CFT on the hemisphere HS$^{d+1}$ by a Weyl rescaling.  
    For a bCFT, the sphere subtraction remains the correct choice, but will require analytic continuation. Geometrically this follows from the intuitive fact that the sphere geometry in embedding space can be obtained from the AdS one by a Wick rotation, together with the continuation $L = i R$. Let us see how this works out precisely:
    suppressing UV log divergences for clarity, a bulk UV counterterm takes the form:
    \be
    y_k \Lambda_{\text{UV}}^{d +1 - 2n - k} \int_{M} \sqrt{g} \cR^n \sim y_k \Lambda_{\text{UV}}^{d + 1 - 2n - k} \frac{ \text{Vol}(M)}{\ell^{2 n}} \, ,
    \ee
    where $y_k$ is a bulk coupling of dimension $k$, $M= \AdS_{d+1}, \, S^{d+1}$ and $\ell = L, R$, respectively. On AdS$_{d+1}$ this object is also IR divergent: holographic renormalization replaces the IR-divergent volume of AdS by its finite regularized $\widetilde{\text{Vol}} (\AdS_{d+1})$ (in even $d$, we take a $\partial_\perp$ derivative to remove the log). Interestingly, the following identity holds:
    \bea
    \kappa_d & =  \frac{\widetilde{\text{Vol}}(\AdS_{d+1})}{\text{Vol}(S^{d+1})} \\ & = \left(\frac{L^2}{R^2} \right)^{(d+1)/2}\begin{cases}
       (-1)^{(d+1)/2} 1/2 \, , \quad & \text{odd $d$} \, , \\
       (-1)^{d/2} 1/\pi \, , \quad & \text{even $d$} \, .
    \end{cases}
    \eea
    Furthermore, the Ricci scalar $\cR$ on the two spaces, which is the only non-vanishing bulk geometric invariant, has equal magnitude and opposite sign. We thus conclude that the combination 
    \be
    \cF_{\AdS} \equiv  (-)^{s_d} \text{Re}\left( F_{\AdS_{d+1}}  - \kappa_d \, F_{S^{d+1}} \bigg\vert_{R = i L} \right) \, .
    \ee
    is free of both bulk UV and IR divergences and thus a physical observable of the bCFT.
    This definition has the advantage of being independent on the choice of reference boundary condition.. The analytic continuation to the sphere may introduce an imaginary part to $F_{S^{d+1}}$, which is automatically removed with the above definition. The need to take the real part can be understood as imposing invariance under the disconnected part of the $O(d+1,1)$ isometry group, on which we will elaborate further later.
\end{enumerate}
To apply procedure 2 consistently, we need a regularization scheme that works for both AdS and sphere geometries. Common options include Pauli-Villars regularization, dimensional regularization and zeta function/heat-kernel methods. A dimensionally regularized BCFT free energy that is monotonic under boundary RG flows has been discussed in several works \cite{Giombi:2014xxa, Giombi:2020rmc, Giombi:2025pxx}. However those results apply only when the bulk theory remains conformal, in which case no analytic continuation of the sphere radius is needed. Our prescription is more general: it applies equally to non-conformal bulk theories, while reducing to the approach of \cite{Giombi:2020rmc} (up to overall normalization and analytic continuation) when the bulk is conformal and dimensional regularization is used.

\section{Boundary Weyl invariance for QFT in AdS}
 To sharpen our discussion of boundary RG flows in AdS and to set the stage for proving the AdS F-theorem in two dimensions, we now discuss how Weyl invariance is realized in bCFTs. We assume the existence of an isometry-preserving boundary condition in AdS and use this to derive constraints on the boundary operator product expansion (OPE) of the bulk stress tensor $T^{\mu\nu}$. Under the same assumptions, we further show that an infinite set of asymptotic charges leaves the boundary condition invariant. This invariance corresponds to Weyl symmetry in bCFT. We then discuss the breaking of AdS Weyl invariance by a boundary RG flow and how to restore it by introducing a boundary dilaton.

 In what follows, we describe the bulk-boundary system through its coupling to the bulk metric $g_{\mu\nu}$ and the embedding function $X^\mu$, which specifies the location of the boundary, following standard treatments in the literature \cite{McAvity:1993ue,Jensen:2015swa,Billo:2016cpy,Cuomo:2021cnb}. 
 We also use the induced metric $h_{ab} = \partial_a X^\mu \partial_b X^\nu g_{\mu\nu}$ as well as the normal vector $n^\mu$. Contraction with the normal vector is denoted by $\perp$, i.e. $n_\mu V^\mu \equiv V^\perp$.
 In AdS, we work in the Poincar\'e coordinates:
 \be
 ds^2 = \frac{dz^2 + \delta_{ab} dx^a dx^b}{z^2} \, .
 \ee
 Employing a finite IR cutoff $z_0$ we have $h_{ab} = \delta_{ab}/z_0^2$. We denote ``rescaled" bCFT observables by a hat, e.g. $\hat{h}_{ab} = \delta_{ab}$.

 \vspace{2mm} \noindent \textbf{Boundary Weyl invariance in AdS.} Let us begin by discussing how the boundary Weyl invariance is realized in a QFT in AdS, following the presentation in Appendix E of \cite{Meineri:2023mps}. 

 An isometry-preserving boundary condition implies that the associated conserved charges
 \be \label{eq: chargeflat}
 Q_\xi = \int_{z=0} \sqrt{h} T^{\mu\nu} n_\nu \xi_\mu \, ,
 \ee
 constructed from a bulk Killing vector $\xi^\mu$, vanish when acting on the boundary. To be concrete, consider inserting $Q_\xi$ on a bulk co-dimension one surface surrounding arbitrary set of bulk operator insertions $X$. Since bulk correlation functions $\langle X\rangle$ are invariant under isometries, shrinking this surface into the bulk gives zero; $\delta_\xi\langle X \rangle = 0$. Because the surface insertion of $Q_{\xi}$
 is topological, the result must remain unchanged as the surface is deformed. In particular, deforming the surface to approach the AdS boundary and evaluating it using the boundary OPE must again give zero.\footnote{One might also consider a more general setup with operator insertions at the boundary, in which case the boundary limit of $Q_\xi$ would need to produce in addition some contact terms localized at the boundary operator insertions.}
 
 Let us discuss the implications for the bOPE. AdS Killing vectors $\xi^\mu$ all have asymptotically finite normal $\xi_\perp = n_\mu \xi^\mu$ and tangential $\xi^a$ components at the AdS boundary. As a result, the associated charge scales near the boundary as
 \be\label{eq:QxiAdS}
 Q_\xi \simAdSzzero  z_0^{-d} \int_{z_0} \sqrt{\hat{h}} T^{\perp \perp} \xi_\perp + z_0^{-d-2} \int_{z_0} \sqrt{\hat{h}} T^{\perp a} \hat{h}_{ab} \xi^b \, .   
 \ee
A spin $l$ bulk field $\Phi^{\mu_1...\mu_l}$ in AdS decomposes into spin $s \leq l$ fields $\Phi^{a_1 ... a_s}$ under the $SO(d) \subset SO(d+1,1)$. Their boundary OPE takes the form
 \be \label{eq: bOPE}
 \Phi(z,x)^{a_1...a_s} \simAdS  b_{\Phi \hat{\cO}} z^{\Delta_{\Hat{\cO}} + |s|}  \Hat{\cO}(x)^{a_1 ... a_s} + ... \, .
 \ee
Applying this scaling to \eqref{eq:QxiAdS}, preservation of AdS isometries imposes stringent constraints on the bOPE of the bulk stress tensor $T^{\mu\nu}$. Scalar primaries in the bOPE must have dimension $\Delta_0 > d$ while vector operators, whether primaries or descendants, must satisfy $\Delta_1 > d +1$. 

Vector operators in the bOPE can arise either as primary spin-1 operators or as descendants of spin-0 or spin-2 operators. Spin-1 and spin-2 primaries appear only in the traceless symmetric component of the bulk stress tensor and are further constrained by the bulk stress-tensor conservation, as in BCFT \cite{McAvity:1993ue,McAvity:1995zd,Lauria:2018klo, Herzog:2017kkj}: only spin 1 primaries of dimension $d+1$ and spin 2 primaries of dimension $>d$ are compatible with conservation. Combining these conditions excludes spin-1 primaries in the bOPE of $T^{\mu\nu}$ altogether, while allowing spin-2 primaries. 

The situation for scalar primaries differs from standard BCFT. In BCFT, the bulk stress tensor is traceless, and together with conservation this fixes the dimension of the scalar primary appearing in the bOPE to be $d+1$. This operator is the familiar {\it displacement operator}. By contrast, in a generic QFT in AdS the bulk stress tensor has a nonvanishing trace, $T = T^\mu{}_\mu \neq 0$. As a result, conservation alone does not constrain the dimension of scalar primaries appearing in the bOPE.\footnote{The conservation law only relates the spin-2 and spin-0 components of $T^{\mu\nu}$ \cite{Ciccone:2024guw}.} Instead, the only restriction comes from the Ward identity associated with AdS isometries discussed above, which requires that scalar operators in the bOPE be irrelevant.

 We denote the leading scalar primary appearing in the bOPE by $\Hat{D}$, with $\Delta_{\Hat{D}} > d$, and refer to it as the “displacement operator.” This terminology follows that used in BCFT; when a QFT in AdS is obtained via conformal perturbation theory around a UV CFT, $\hat{D}$ may be viewed as a natural continuation of the BCFT displacement operator.  In this sense, although its scaling dimension is unprotected, the operator retains a clear physical interpretation as governing the response to boundary deformations.

Note that the boundary OPE constraints discussed above are not automatically satisfied by an arbitrary choice of conserved bulk stress tensor. Rather, they provide a criterion for selecting the isometry-preserving stress tensor among the possible choices related by improvement terms. 
This need for improvement is physically natural. When a UV CFT in AdS is perturbed by a relevant operator, there is no unique prescription for coupling the perturbation to the curved background. As a simple example, consider a free massive scalar: at the level of the action, a bare mass term is indistinguishable from a conformal mass, yet the two lead to different bulk stress tensors. The correct form of $T^{\mu\nu}$ is fixed by requiring preservation of AdS isometries. This can be achieved through an improvement of the form\footnote{Note that in general there can also be a nonzero VEV of the bulk stress tensor, that generates an expectation value for the charges $Q_\xi$. Such contribution can always be subtracted using an improvement proportional to the bulk metric times the bulk identity operator. It can also be seen as the special case of \eqref{eq: improvement} in which $\Phi$ is the bulk identity operator. In perturbative calculations, e.g. for the stress tensor of free fields, this subtraction is naturally implemented by normal ordering.}
\be \label{eq: improvement}
\delta T_{\mu\nu} = \left( G_{\mu\nu} \Box - \nabla_\mu \nabla_\nu + \cR_{\mu\nu} \right) \Phi(X) \, .
\ee
where $\Phi$ is an arbitrary bulk scalar field.
$\delta T_{\mu\nu}$ is identically conserved and shares the same scalar conformal blocks as the bulk stress tensor. It can therefore be used to remove unwanted relevant scalar contributions from the boundary OPE of $T^{\mu\nu}$.  If we have $n$ relevant scalars in the OPE instead, the bulk fields $\Phi^{(j)} = \Box^j T$, $j=0, ..., n-1$ give a linearly independent basis for the subtraction.

As we adiabatically change $\Lambda L$, the displacement operator can reach marginality. In this case the AdS isometries are explicitly broken as pointed out in \cite{Copetti:2023sya,Lauria:2023uca} and the bCFT is destabilized and undergoes a boundary RG flow.

We now describe boundary Weyl transformations. Unlike the BCFT case, where boundary Weyl rescalings descend from bulk Weyl symmetry, a bCFT is coupled to a massive bulk theory that is not Weyl invariant.
Instead, we make use of the fact that the boundary of AdS must be understood in a {\it conformal} sense; only the Weyl equivalence class of its induced metric $\hat{h}_{ab}$ is well-defined.
This equivalence is generated by a class of bulk diffeomorphisms known as Penrose–Brown–Henneaux (PBH) transformations \cite{Penrose:1985bww,Brown:1986nw,Imbimbo:1999bj}. These transformations preserve the asymptotic AdS metric up to subleading terms near the boundary while inducing a Weyl rescaling of the boundary metric.\footnote{More precisely, $\delta_\xi g_{\mu\nu} =\delta_\mu^a \delta_\nu^b\left( 2 \tau\,  \hat{h}_{ab} + \hat{\nabla}_a \chi_b + \hat{\nabla}_b \chi_a \right)/z^2 + O(1/z)$. The term in parenthesis denotes a combination of a boundary Weyl transformation $\tau$ and diffeomorphism $\chi^a$.} 

In Poincar\'e coordinates the PBH vector fields take the asymptotic form
\be \label{eq: xiPBH}
\xi^\mu_{\text{PBH}} = \left(-  z \tau(x), \, z \frac{1}{2}\partial^a \tau(x) + \chi^a(x) \right) \, .
\ee
Here $\tau$ describes a boundary Weyl rescaling, while $\chi^a$ acts as a diffeomorphism on the boundary coordinates. 
Importantly, the PBH transformations also satisfy that both $\xi_\perp$ and $\xi^a$ are finite near the AdS boundary. Thus, the structure of the bOPE discussed above guarantees that, if the boundary condition preserves the AdS isometries (i.e.~ $Q_{\xi_{\rm isometry}}=0$), it is also invariant under the entire family of asymptotic charges:
\be
Q_{\xi_{\text{PBH}}} = 0 \, .
\ee
This shows that AdS isometries of a bCFT are enlarged to a larger boundary Weyl group, establishing the boundary Weyl invariance of bCFTs.

\vspace{2mm} \noindent \textbf{Boundary deformations and the dilaton.}
We now consider the breaking of AdS Weyl invariance by a boundary RG flow, and how to remedy this by the coupling to a boundary dilaton $\omega$. Consider a bulk scalar field $\Phi(z,x)$ with boundary OPE \eqref{eq: bOPE} (with $s=0$). With a finite IR regulator $z_0$ we define the boundary deformation:
\be \label{eq:deform}
\delta S_{\text{bdry}} = \lambda z_0^{d-\Delta_{\hat{\cO}}} \int \sqrt{h} \Phi(z_0,x) = \lambda  \int \sqrt{\hat{h}} \hat{\cO}(x) + ...~ ,
\ee
where we have chosen a normalization of $\Phi$ in which $b_{\Phi \widehat{\cO}} = 1$.
The parameter $\lambda$ can be thought of as a function $\lambda(M z_0)$, where $M$ is a mass scale controlling the boundary RG. 
Due to the explicit cutoff dependence, this deformation breaks the AdS isometries. 

A boundary RG transformation corresponds to changing $M$, or equivalently $z_0$, while leaving the boundary metric fixed. This is achieved by a Weyl transformation accompanied by a compensating diffeomorphism \cite{Baume:2014rla}. This transformation is induced by a PBH vector field $\xi^\mu_{\text{RG}}$ whose components satisfy 
\begin{equation}
\xi^\mu_{\text{RG}}:~~\widehat{\nabla}_a \chi_b + \widehat{\nabla}_b \chi_a = - 2 \tau \hat{h}_{a b}~,
\end{equation}
which includes the case of bulk isometries. Note that the Weyl rescaling corresponds to varying $z_0$ along a space-dependent profile. The resulting charge acting on the boundary gives
\begin{align}
\begin{split}\label{eq: breaking}
\lim_{z\to 0} Q_{\xi_{\text{RG}}} & \equiv \int \sqrt{\hat{h}} \,\tau(x)\, \widehat{\Theta}(x) \\
& = \beta_\lambda \int \sqrt{\hat{h}} \,\tau(x) \,\hat{\cO}(x) \, ,
\end{split}
\end{align}
where $\beta_\lambda = M \frac{d}{d M}\lambda $. The existence of a nontrivial boundary operator $\widehat{\Theta}(x)$ that controls the boundary beta functions, as defined by \eqref{eq: breaking},  implies that the leading behavior of the components $T^\perp_{~\perp}$ and $T^\perp_{~a}$ of the stress tensor is $\sim z_0^d$ (with unbroken AdS isometries we saw above that the behavior of both components is further suppressed).\footnote{In order to compute $\widehat{\Theta}(x)$ one needs to use that, even in the presence of a boundary RG, boundary isometries are still a symmetry. This implies that $T^\perp_{ ~~ a} \underset{z_0\to 0}{\sim} z_0^{d} \,\widehat{\nabla}_b \widehat{\tau}^b_{~~a}$, for symmetric $\hat{\tau}^{ab}$. Defining $T^\perp_{~\perp} \underset{z_0\to 0}{\sim} z_0^{d} \,\widehat{\Sigma}$, we have $\widehat{\Theta} = -(\widehat{\Sigma} + \hat{h}^{ab} \widehat{\tau}_{ab})$.\label{foot:hattheta}}

\begin{figure}
    \centering
   \begin{tikzpicture}
  \newlength\R
  \setlength\R{1.5cm}

  \begin{scope}
    \clip (0,0) circle (\R);
    \node[anchor=center,inner sep=0] at (0,0)
      {\includegraphics[width=\dimexpr2\R\relax]{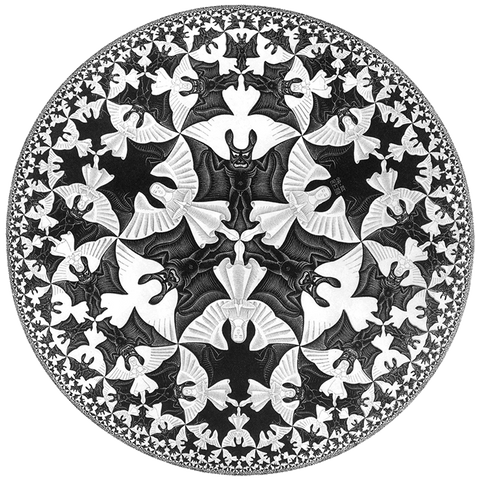}};
  \end{scope}

  \draw[line width= 3] (0,0) circle (\R);
\node at (0,-2) {(a)};
   
  \begin{scope}[shift={(4,-0.12)}]
 \node at (0,-1.88) {(b)};
    \clip[decorate,decoration={snake, amplitude=2pt, segment length=15pt}]
         (0,0) circle (\R);
    \node at (0,0.12)
      {\includegraphics[width=\dimexpr2.2\R\relax]{Figures/escher.png}};

  \draw[line width=4,
        decorate,decoration={snake, amplitude=2pt, segment length=15pt}]
        (0,0) circle (\R);
       
  \end{scope}
\end{tikzpicture}
    \caption{(a) Global Euclidean AdS space. (b) Introducing a modulated dilaton on the AdS boundary by a PBH transformation.}
    \label{fig:adsdilaton}
\end{figure}

We can restore Weyl invariance by explicitly coupling the system to a boundary dilaton $\omega$. This can be done via a modulated embedding (see Figure \ref{fig:adsdilaton})
$X^\mu = (z_0 e^{\omega(x)}, x^a)$.
An analogous soft breaking of AdS isometries has been used to great effect in the context of Jackiw-Teitelboim gravity \cite{Maldacena:2016upp,Jensen:2016pah}.
While this restores Weyl invariance, the free energy now depends explicitly on the dilaton profile, which appears through the combination
 $M e^{\omega}$. This trick of restoring Weyl invariance by a background dilaton has been very effective for proving monotonicity theorems \cite{Komargodski:2011vj,Komargodski:2011xv,Cuomo:2021rkm}, and we will similarly make use of it in this Letter.

\vspace{2mm} \noindent \textbf{Example: free massive scalar in AdS.}
\begin{figure}[t]
    \centering
    \includegraphics[width=0.99\linewidth]{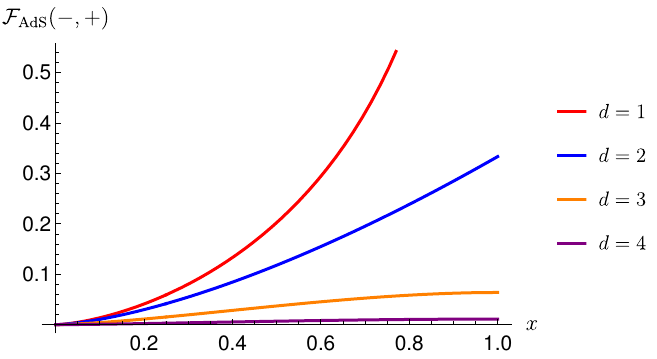}
    \caption{Free energy difference between $+$ and $-$ b.c. as a function of $x=\tfrac{4M^2L^2+d^2}{\text{Min}\left[d^2,4\right]}\in[0,1]$, parametrizing the range of masses in which both $\pm$ b.c. are consistent.}
    \label{fig:freescalarprl}
\end{figure}
Since the discussion above has been rather abstract, we will now analyze the concrete example of a free massive scalar field $\phi$ of mass $m$ in AdS$_{d+1}$ to showcase the salient features of the AdS free energy and the breaking of boundary Weyl invariance. For masses in the range $-d^2/4 \leq m^2 L^2 \leq \min(0,1-d^2/4)$, the theory has two simple AdS-isometric boundary conditions, which we denote by $\pm$. The corresponding bCFTs are the mean field theories of generalized free fields $\hat{\cO}_\pm$ with dimension $\Delta_\pm = \frac{d}{2} \pm \sqrt{\frac{d^2}{4} + m^2 L^2} \equiv \frac{d}{2} \pm \nu$.
The $-$ b.c.~is unstable, and flows to the $+$ b.c.~by a double-trace deformation $\frac{f}{2}\int \hat{\cO}_-^2$.\footnote{The double trace flow has been studied in several works both in the context of AdS/CFT \cite{Aharony:2001pa,Gubser:2002vv,Gubser:2002zh,Hartman:2006dy,Diaz:2007an} and BCFT \cite{Jensen:2015swa,Herzog:2019bom,Wang:2020xkc,Chalabi:2022qit}.}
It can be shown (see Supplemental Material) that, upon judiciously fixing the improvement ambiguity \eqref{eq: improvement}, the boundary asymptotics of the stress tensor with $-$ b.c.~are $T^{\perp \perp}(z_0,x) \simAdSzzero -\frac{1}{2 (2 \Delta_- + 1)} z_0^{d + 2 - 2\nu} \left[ \hat{\partial}_a \hat{\cO}_- \hat{\partial}^a \hat{\cO}_- - \frac{\Delta_-}{2 \Delta_- - d +2} \hat{\cO}_- \widehatBox \hat{\cO}_- \right]$, $T^{\perp a}(z_0,x) \simAdSzzero \mathcal{O}(z_0^{d +1+ 2(1-\nu)})$, in agreement with the constraints from the AdS isometries. 
We identify the operator 
\be
\hat{D}=-\frac{1}{2 (2 \Delta_- +1)}\left[ \hat{\partial}_a \hat{\cO}_- \hat{\partial}^a \hat{\cO}_- - \frac{\Delta_-}{2 \Delta_- - d +2} \hat{\cO}_- \widehatBox \hat{\cO}_- \right]
\ee
as the ``displacement''. Indeed this is the standard displacement operator for Neumann boundary conditions when $\nu=1/2$. Notice that this is irrelevant for the range of boundary conditions we are considering. Once this becomes relevant, the $-$ b.c. ceases to exist. 

On the other hand, in the presence of the double trace deformation the boundary conditions become \cite{Klebanov:1999tb,Aharony:2015afa} $\hat{\cO}_+ = -\frac{f}{2\nu}  \hat{\cO}_-$. As per the discussion above, in this case the components of the stress tensor receive new terms with asymptotic behavior $\sim z_0^{d}$ caused by the breaking of the isometries. Using these modified asymptotics we find
\be
\widehat{\Theta} = (d - 2 \Delta_-) \frac{f}{2} \hat{\cO}_-^2~.
\ee
We recognize the classical beta function of the boundary coupling $f$.

Let us now examine the AdS free energy $F_{\AdS}(\pm)$.\footnote{The calculation of these partition functions, and a check of their monotonicity, has been performed, with a different technique, already in \cite{Giombi:2014xxa}, however with a different interpretation, namely as the monotonic partition function of a {\it local and holographic} boundary theory in the limit of large $N$.} It satisfies:
\begin{align}
\begin{split}
& \frac{\partial}{\partial m^2 L^2} F_{\AdS}(\pm) =  \frac{\text{Vol}(\AdS_{d+1})}{L^{d+1}} \frac{\Gamma(\frac{d}{2})}{8 \pi^{d/2+1} }\times \\
& \int_{\cC^\pm} d \nu \frac{1}{\nu^2 + \frac{d^2}{4} + m^2 L^2} \frac{\Gamma(\frac{d}{2} + i \nu)\Gamma(\frac{d}{2} - i \nu)}{ \Gamma( i \nu) \Gamma(-i \nu)} \, .
\end{split}
\end{align}
The integral is both UV and IR divergent. The IR divergence is resolved by substituting the AdS volume with its regulated version $\widetilde{\text{Vol}}(\AdS_{d+1})$. After subtracting-off the sphere contribution, we obtain a  finite scheme-independent answer for the derivative of the free energy, as shown in Table \ref{tab:freescalar}.

\begin{table}[t]
    \centering
    \begin{tabular}{|c|c|} \hline
       even $d$  & $\ds \frac{\partial}{\partial \nu} \cF_{\AdS} =  \frac{(-1)^{d/2}\nu \Gamma(\frac{d}{2}+\nu)}{ \Gamma(d+1) \Gamma(1-\frac{d}{2} + \nu)}$ \\ \hline 
       $d=2$ & $\ds- \frac{1}{2}\nu^2$ \\ \hline 
       $d=4$  & $\ds-\frac{1}{24} \nu^2 (\nu^2 - 1)$  \\ \hline\hline
       odd $d$ & $\ds \frac{\partial}{\partial \nu} \cF_{\AdS} =  \frac{(-1)^{(d+1)/2}\pi \nu\Gamma(\frac{d}{2}+\nu)}{2 \Gamma(d+1) \Gamma(1-\frac{d}{2}+\nu)} \tan(\pi\nu)$  \\ \hline 
       $d=1$ &  $\ds - \frac{\pi}{2} \nu \tan(\pi \nu)$ \\\hline 
       $d=3$  &   $\ds- \frac{\pi}{48} \nu (1 - 4 \nu^2) \tan(\pi \nu)$ \\ \hline
    \end{tabular}
    \caption{$\frac{\partial}{\partial \nu}  \cF_{\AdS}$ for the free massive scalar. The free energy with $\pm$ boundary conditions is given by:
    $\mathcal{F}_{\text{AdS}}(+)=\int_0^\nu\!d\nu' \tfrac{\partial}{\partial\nu'}\mathcal{F}_{\text{AdS}}= \left.\mathcal{F}_{\text{AdS}}(-)\right|_{\nu\rightarrow-\nu}$. }
    \label{tab:freescalar}
\end{table}
The free energy can be obtained by integrating over $m^2 L^2$. The integration constant can either be fixed by matching to the massless scalar BCFT answer, which corresponds to the conformal mass $m^2 L^2 = - (d^2-1)/4$, or by requiring that the $+$ and $-$ free energies coincide at $m^2 L^2 = -d^2/4$. The integrated form is not illuminating, but we can explicitly check the monotonicity statement: $\cF(-) - \cF(+) \geq 0$ in the full allowed range of parameters. Selected plots are shown in Figure \ref{fig:freescalarprl}.

As a sanity check let us consider the case of $d=2$ explicitly. The $\pm$ b.c. merge at $m^2 L^2 = -1$, while at $m^2 L^2 = - 3/4$ we should recover the conformal scalar, for which $\cF_{\text{conf.}}(-,+) = \Delta b/3 =  1/24$, where $b$ is the b-function \cite{Jensen:2015swa}. Integrating the derivative of the free energy we find $\cF_{\AdS}(-,+) = \frac{1}{3} (1 + m^2 L^2)^{3/2} $, which matches expectations for both known values of the mass.
For bCFT$_3$, it would be interesting to understand the connection between our AdS free energy and boundary Weyl anomalies, perhaps generalizing the b-theorem to boundaries of massive QFTs.
The example of a free, massive fermion can be treated similarly, and can be found in the Supplemental Material.

\section{Proof in AdS$_2$}
We now give a proof of the AdS F-theorem in $d=2$. This generalizes the known discussion of the g-theorem in BCFT$_2$ \cite{Affleck:1991tk,Friedan:2003yc,Cuomo:2021rkm}. 
In the BCFT setup, the breaking of boundary conformal symmetry is compensated by the introduction of a dilaton field $\omega$. The salient step in these proofs is to show that the dilaton effective action possesses a non-linearly realized symmetry:
\be \label{eq:nonlinward}
\omega \longrightarrow \omega + \varepsilon( \dot{\xi}^\alpha +  \xi^\alpha \dot{\omega}) \, ,
\ee
where $\xi^\mu$ is a bulk conformal killing vector. This symmetry is derived by explicitly computing the action of the (bulk) conformal isometry charge on the boundary degrees of freedom.
As the dilaton couples linearly to the trace of the boundary stress tensor $\Theta$, the Ward identities stemming from \eqref{eq:nonlinward} amount to nontrivial constraints relating the integrated 1- and 2-point functions of $\Theta$. A careful manipulation of these identities leads to the monotonicity constraint.

In global coordinates the AdS$_2$ metric is $ds^2 = d\eta^2 + \sinh^2(\eta) d \alpha^2$, with $\alpha \in [0,2\pi)$ parametrizing the boundary $S^1$. The radius of the $S^1$, which we denote by $\ell$ only becomes physical after breaking the AdS$_2$ isometry, and gives a length scale for the boundary RG. It can be understood as a background (constant) value for the dilaton field.
The AdS isometries form the algebra $\mathfrak{sl}(2)$ with generators:
\bea
&\xi_0 &&= - \partial_\alpha \, , \\
&\xi_{1} &&= -\frac{\cos(\alpha)}{\tanh(\eta)} \partial_\alpha - \sin(\alpha) \partial_\eta  \\
&\xi_{-1} &&= - \frac{\sin(\alpha)}{\tanh(\eta)} \partial_\alpha + \cos(\alpha) \partial_\eta 
\eea
In the presence of a bCFT deformation \eqref{eq:deform}
these are broken to translations $\xi_0$. We amend this by coupling to a boundary dilaton as in Figure \ref{fig:adsdilaton}.
As in \cite{Cuomo:2021rkm}, an AdS isometry now relates two different dilaton backgrounds, giving rise to a symmetry of the dilaton effective action. The relation is encoded in the charge $Q_\xi = \int \sqrt{h} T^{\mu\nu} \xi_\mu n_\nu $, which is topological in the bulk, and can be trivialized at the origin of AdS, regardless of the boundary RG.\footnote{Notice that these are the only PBH transformations for which this argument can be applied, as it relies on our ability to deform the symmetry operator into the bulk of AdS: asymptotic symmetries only correspond to approximate topological operators near the boundary.} By virtue of $X^\mu = (\eta_0 + \omega(\alpha), \, \alpha)$, $n_\mu = (-1, \dot{\omega}(\alpha) + O(e^{-\eta_0}))$ and $\hat{\Gamma}_{\alpha \alpha}^\alpha = \dot{\omega}(\alpha)$ the expression for the charge simplifies to:
\be
Q_\xi = \int \sqrt{\hat{h}} \widehat{\Theta} (\dot{\xi^\alpha} + \dot{\omega} \xi^\alpha)\, ,
\ee
where the asymptotic identity $\xi^\eta \simAdS -\dot{\xi}^\alpha$ was also used. This charge is finite for a boundary RG flow. As $\widehat{\Theta}$ couples linearly to the dilaton we conclude that the free energy is invariant under the linearized transformation:
\be
\omega \longrightarrow \omega +  \varepsilon\left( \dot{\xi}^\alpha + \xi^\alpha \dot{\omega} \right) .
\ee
Notice that this is parallel to the BCFT setup.

The considerations above imply that $\langle Q_\xi \rangle_\omega = 0 $, where $\langle \cdot \rangle_\omega$ denotes the expectation value in the presence of a background dilaton $\omega$. This holds regardless of the chosen profile for $\omega$. Expanding this identity to first order for small $\omega$ we obtain
\bea
&\int d \alpha \sqrt{\hat{h}} \xi^\alpha(\alpha)  \dot{\omega(\alpha)} \, \langle \widehat{\Theta} \rangle = \\ 
&- \int d \alpha_1 \sqrt{\hat{h}_1} \int d\alpha_2 \sqrt{\hat{h}_2} \dot{\xi}^\alpha(\alpha_1) \, \omega(\alpha_2) \langle \widehat{\Theta}(\alpha_1) \widehat{\Theta}(\alpha_2) \rangle \, . 
\eea
Due to the broken AdS isometries, the AdS$_2$ free energy suffers from \emph{boundary} UV divergences, stemming from the OPE of the perturbing operator. In $d=1$, there are only two relevant counterterms by power counting: the boundary cosmological constant $\Lambda_{\text{UV}}\,\int \sqrt{\hat{h}}$ and the finite extrinsic curvature counterterm $\int \sqrt{\hat{h}} K$. The former is linearly dependent on $\ell$ and can be subtracted by considering the boundary entropy:
\be
\cS_{\AdS}(M_0 \ell e^{\omega}) = \left(1 - \ell \frac{\partial}{\partial \ell}\right) \cF_{\AdS} \, .
\ee
Let us now discuss why the finite counterterm does not contribute. Notice that, with our definition in \eqref{eq:Fads}, before turning on any boundary deformations $\cF_{\AdS}$ is invariant under the full $O(2,1)$ symmetry group. This includes an anti-unitary CPT symmetry which reverses the direction of the AdS normal vector by exchanging the two sheets of the hyperboloid $-X_0^2 + X_1^2 + X_2^2 = - L^2$.  The boundary relevant deformation preserves this discrete symmetry. The finite counterterm is odd under this symmetry and must thus have an imaginary coefficient, therefore dropping out of the free energy. 
Finally, following \cite{Cuomo:2021rkm}, we find
\bea
&M \frac{\partial}{\partial M} \cS_{\AdS} =
\\ &- \int d \alpha_1 \sqrt{\hat{h}_1} \int d\alpha_2 \sqrt{\hat{h}_2} \langle \widehat{\Theta}(\alpha_1) \widehat{\Theta}(\alpha_2) \rangle (1- \cos(\alpha_1 - \alpha_2)) \, ,
\eea
which is manifestly negative definite and finite due to the double zero of the cosine term. This concludes the proof of the F-theorem in AdS$_2$.

Let us comment briefly on the higher-dimensional generalizations. First, it is clear that the dilaton effective action in AdS can be generalized to arbitrary dimensions. This might lead to a proof of the AdS F-theorem in $d=3$ --generalizing the b-theorem in BCFT \cite{Jensen:2015swa}-- once the role of boundary Weyl anomalies in AdS is clarified. 
For boundary RG flows controlled by one-loop conformal perturbation theory (e.g. short flows), the methods of \cite{Klebanov:2011gs} generalize readily to the bCFT case, as the integrated correlation functions computing the free energy in conformal perturbation theory are similarly fixed by conformal symmetry. Finally, it is certainly promising to extend entanglement-based methods of proof to cover the AdS case, building on \cite{Casini:2018nym,Casini:2022bsu,Casini:2023kyj}. We are aware of current efforts in this direction by \cite{entanglement}.

\section{Applications}
We conclude with two applications of our results. The first concerns RG flows between non-local CFT$_d$s, which can be described by free massive scalars in AdS$_{d+1}$ coupled to boundary degrees of freedom. The AdS F-theorem gives a natural generalization of the standard F-theorem to the non-local case. 
The second consists of exactly solvable interacting systems in AdS$_2$, namely the $O(N)$ sigma model and the Gross-Neveu model. These are toy models for confinement, as they exhibit dynamical mass generation. The AdS F-theorem corroborates the results of \cite{Copetti:2023sya} about the structure of boundary phase transitions in these models.

\subsection{Long-range CFTs}
Long-range CFTs reached by a relevant deformation of a Gaussian Generalized-Free-Field (GFF$_\Delta$, where $\Delta$ is the scaling dimension of the GFF) have a natural description in terms of massive free fields in AdS, whose $\pm$ boundary conditions describe the GFF$_{\Delta_\pm}$ dynamics. Due to their nonlocal nature, these theories, if seen as d-dimensional QFTs, do not admit a local conserved stress tensor. Thus, the standard arguments for the monotonicity of their free energy do not apply straightforwardly. If instead we interpret these theories as boundary conditions for QFTs in AdS, our $\cF_{\text{AdS}}$ theorem provides a natural setting for studying their RG flows.

\vspace{2mm}\noindent \textbf{Long-Range Ising CFT.}  A simple example of this correspondence is the Long-Range Ising (LRI) CFT. Let us briefly review two possible realizations of the LRI CFT \cite{Behan:2017emf, Behan:2017dwr}. It is useful to introduce the variable $s$ defined by $\Delta_\pm = \frac{1}{2}(d\pm s)$, restricted to $s\leq\text{min}(d,2)$ to ensure that $\Delta_-$ is above the unitarity bound. The first realization is obtained perturbing the Gaussian theory of the scalar  GFF$_{\Delta_-}$ $\phi(x)$ via the quartic interaction
\be\label{def-}
S_{\text{int}} = \frac{\lambda}{4} \int d^d x \, \phi(x)^4 \, .
\ee
For a range of values of $s$ with $s\geq s_- \equiv d/2$ this deformation is relevant and the IR fixed point is the LRI CFT. The second description is obtained starting with the product of the Gaussian theory of the scalar GFF$_{\Delta_+}$ $\tilde{\phi}(x)$ with the $d$-dimensional (local) Ising CFT, and deforming it by the following interaction between the two sectors
\be\label{def+}
S_{\text{int}} = \widetilde{\lambda} \int d^d x \, \widetilde{\phi}(x) \sigma(x) \, , 
\ee
 where $\sigma(x)$ is the energy operator of the Ising CFT. For a range of values of $s$ with $s \leq s_+ \equiv d - 2\Delta_\sigma$ this deformation is relevant and the IR fixed point is the LRI CFT. Putting the two descriptions together, for $s_- \leq s \leq s_+$ both RG flows have the LRI as IR fixed point. Moreover, for $s> s_+$ the IR fixed point of \eqref{def-} is GFF$_{\Delta_+}\otimes\,$Ising, and for $s< s_-$ the IR fixed point of \eqref{def+} is GFF$_{\Delta_-}$.

\begin{figure}[t]
  \includegraphics[width=0.95\linewidth]{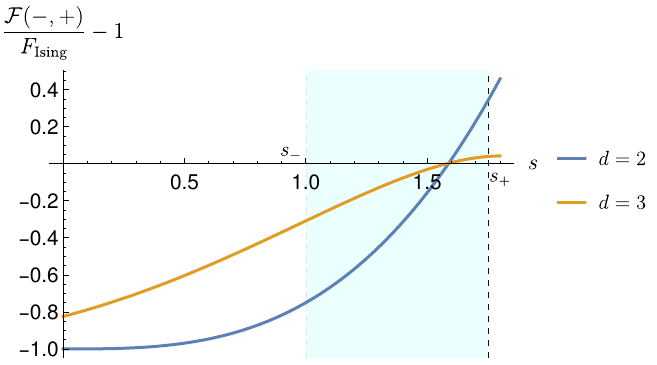}
    \caption{The AdS F theorem for the LRI CFT. Dashed vertical lines mark the transition points between GFF-LRI and SRI-LRI, the LRI phase is the stable RG fixed point in the shaded region. We have rescaled $s_{d=3}$ so as to make the dashed transition lines coincide between $d=2$ and $d=3$. By the AdS F-theorem and \eqref{eq: ineq} the plotted function must be negative for $s< s_-$ and positive for $s>s_+$. For the $d=3$ plots we have used the numerical estimates $\Delta_\sigma \sim 0.5181$ \cite{Chang:2024whx} and $F_{\text{Ising}} = 0.0612$ \cite{Hu:2024pen}. We restrict to $s\leq\min(d,2)$ to ensure unitarity of the $-$ boundary condition.}
    \label{fig:longrange}
\end{figure}

\begin{figure}[t]
    \includegraphics[width=0.95\linewidth]{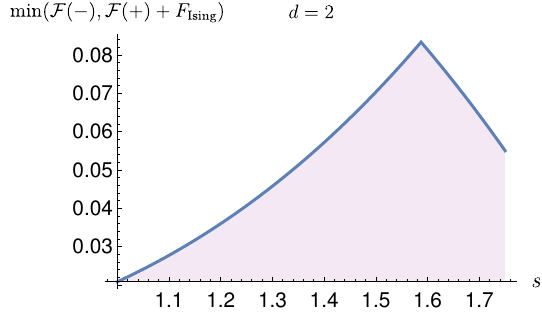}
    \\[0.5em]
    \includegraphics[width=0.95\linewidth]{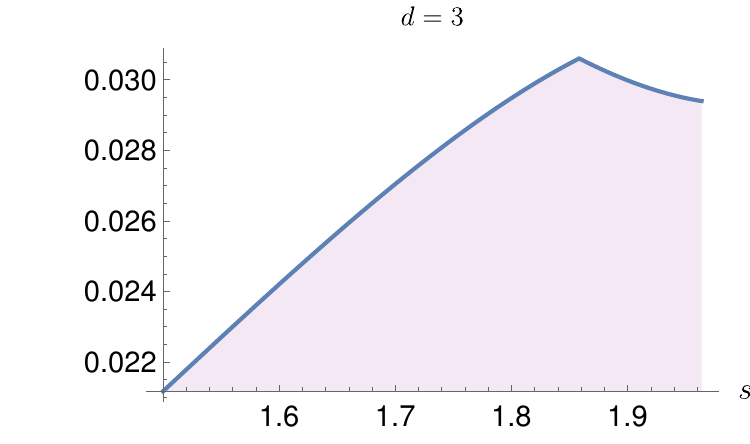}
    \caption{Upper bounds on the LRI free energy in $d=2$ and $d=3$ for $s_-<s<s_+$. The shaded purple regions are allowed by the AdS F-theorem.}
    \label{fig:longrangebounds}
\end{figure}

The AdS F-theorem can be applied to this setup simply by describing the GFF theory as a free bulk field with mass $m^2L^2 = \Delta (d- \Delta)$. This leads to the following inequalities
\renewcommand{\arraystretch}{1.3}
\be \label{eq: ineq}
\begin{array}{ll}
 \cF(-) \leq \cF(+) + F_{\text{Ising}} ,    &0 \leq s \leq s_- \, , \\  
 \cF_{\text{LRI}} \leq \min(\cF(-), \, \cF(+) + F_{\text{Ising}}),    &s_- \leq s \leq s_+ \, \ \\
 \cF(+) + F_{\text{Ising}} \leq \cF(-),   & 
 s \geq s_+ \, .
\end{array}
\ee
Focusing on $d=2,3$ it is straightforward to make use of our previous results for the massive scalar free energies $\cF(\pm)$ to check the validity of the inequalities $\cF(-,+) < F_{\text{Ising}}$ and $\cF(-,+)> F_{\text{Ising}}$ for small and large $s$ respectively (Figure \ref{fig:longrange}).\footnote{Notice an interesting feature: after the $s$ parameters are rescaled to make the known transition lines coincide, the intersection points between the two curves and the horizontal axis look indistinguishably close to each other on the plot. Using the state-of-the-art precisions of $F_\text{Ising}$ and $\Delta_\sigma$ in the literature, we find that the intersecting regions of the three lines still have a non-zero overlap.} Furthermore, the middle inequality can be used to obtain an upper-bound for $\cF_{\text{LRI}}$ (Figure \ref{fig:longrangebounds}).


\subsection{Large $N$ models}
In \cite{Copetti:2023sya}, using the results of \cite{Carmi:2018qzm}, the problem of dynamical mass generation in the large $N$ $O(N)$ and Gross Neveu models was studied in AdS$_2$. In both cases a symmetry-breaking boundary condition is stable for small $\Lambda L$, but cannot reach the flat space regime $\Lambda L \gg1$ due to a boundary phase transition. We show that this picture is consistent with the AdS F-theorem. We refer the reader to \cite{Carmi:2018qzm,Copetti:2023sya} for details.

\begin{figure}[t]
    \centering
     \includegraphics[width=0.95\linewidth]{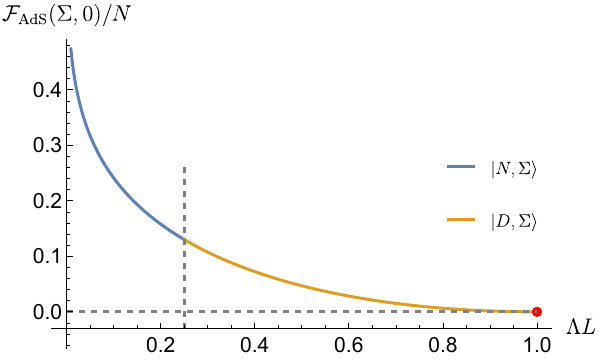}
    \caption{The large $N$ free energy difference between symmetry-preserving and symmetry-breaking b.c. for the O(N) model.}
    \label{fig:ON}
 \end{figure}

\vspace{2mm} \noindent \textbf{$O(N)$ model.} The $O(N)$ model 
\bea
S = \int d^2x \sqrt{g}\left( \frac{1}{2} (\partial \phi)^2 + \sigma \left( \phi^2 - \frac{N}{g^2}\right)  \right) 
\\+ \frac{N}{2} \Tr \log(-\Box + 2 \sigma) \, ,
\eea
in AdS supports two simple families of boundary conditions:
\begin{enumerate}
    \item Massless (symmetry-breaking) Dirichlet boundary conditions $|\text{D}, \Phi^2\rangle$ characterized by a nontrivial vev for the $O(N)$ field:
    \be
   N^{-1} \langle \phi^2 \rangle = \Phi^2 = - \frac{1}{2\pi} \log (\Lambda L) \, ,
    \ee
    with $\Lambda =\mu e^{- 2 \pi/g^2}$ the dynamically generated scale. These boundary conditions come in a continuous family parametrized by $\Phi$ and they are favored at small $\Lambda L$. They describe an AdS spontaneous symmetry-breaking phase which is forbidden in flat space. A continuous transition to a symmetry preserving boundary condition happens at intermediate coupling, consistently with symmetry restoration in the flat-space limit.
    \item Massive (symmetry-preserving) boundary conditions $|\text{D}, \Sigma\rangle$ and $|\text{N}, \Sigma \rangle$, where the Hubbard-Stratonovich vev $\Sigma$ satisfies the gap equation:
    \be
    \frac{1}{g^2} = \text{Vol}(\AdS_2)^{-1} \Tr \left( \frac{1}{\Box - 2 \Sigma} \right)_{\text{D}/\text{N}} \, .
    \ee
    The $|\text{N},\Sigma\rangle$ boundary condition exists only in the weak coupling region, smoothly continuing into $|\text{D},\Sigma\rangle$ at $\Sigma=-1/8$. This is the boundary condition that extrapolates continuously to flat space.
\end{enumerate}
The free energy of the $|\text{D}, \, \Phi^2 \rangle$ b.c. is $\Lambda L$-independent at the leading order in the large $N$ limit, and it coincides with the one of the massless symmetry-preserving b.c. $|\text{D}, \Sigma=0\rangle$. Furthermore, the N and D b.c. merge at the BF bound. The derivative of:
\be
F(\Sigma) = - N \left[ \widetilde{\text{Vol}}(\text{AdS}_2) \frac{\Sigma}{g^2} - \frac{1}{2}\Tr\log(-\Box + 2 \Sigma)_{\text{D}/\text{N}} \right] \, ,
\ee
can be computed from the free scalar results. By the matching conditions above we can integrate the AdS free energies and plot their difference $\cF_{\text{AdS}}(\Sigma,0)= \cF(\Sigma) - \cF(0)$ in figure \ref{fig:ON}. The result for the difference is compatible with the existence of an RG flow from the symmetry-preserving to the symmetry breaking b.c., induced by the relevant leading $O(N)$ vector in the bOPE of $\phi^i$. The difference becomes zero when the symmetry-breaking b.c. coincides with the symmetry-preserving one, consistently with the predictions of \cite{Copetti:2023sya} that it stops existing past that point. 

\vspace{2mm} \noindent \textbf{Gross-Neveu model.}
The Gross-Neveu model:
\bea
S = \int d^2x \sqrt{g}\bigg( \Bar{\psi}^i \slashed{\nabla} \psi_i &+ \sigma \Bar{\psi}^i \psi_i - \frac{N}{2 g} \sigma^2 \bigg) \\ 
&- N \Tr \log\left( \slashed{\nabla} - \sigma \right) \, ,
\eea
also admits two different families of boundary conditions:
\begin{enumerate}
    \item Axial preserving boundary conditions $|A, \eta \rangle$, which are massless ($\Sigma=0$) and break the $O(2N)$ symmetry down to $O(N)$. These b.c. exist at weak-coupling but do not describe the flat-space physics.
    \item Vector-preserving boundary conditions $|V,\Sigma\rangle$, which break the $\bZ_4^A$ axial symmetry down to its $\bZ_2^F$ fermion parity subgroup. They support a non-zero fermion mass, described by a vev $\Sigma$ for the Hubbard-Stratonovich field solving the gap equation:
    \be
    \frac{\Sigma}{g} - \text{Vol}(\AdS_2)^{-1}\Tr \left(\frac{1}{\slashed{\nabla} + \Sigma} \right) = 0 \, .
    \ee
    As the $\bZ_4^A$ is broken, it maps solutions with $\pm \Sigma$ into each other. These b.c. always form a doublet and smoothly interpolate to the flat-space regime.  
\end{enumerate}
The free energy $F_M$ on $M=\AdS_2, S^2$ is given by:
\be
F_M(\Sigma) = - N\left[ \text{Vol}(M)\frac{\Sigma_M^2}{2 g} - F_M^{\text{free}-\psi}(\Sigma_M) \right] \, ,
\ee
where $\Sigma_M$ is the solution to the gap equation and $F_M^{\text{free}-\psi}$ the free energy of a free massive fermion.
It can be checked that $\Sigma \to 0$ at small AdS length. Thus in the weakly coupled regime the two free energies should match the g-function of the free Dirac fermion CFT's boundary conditions \cite{Affleck:1991tk}\footnote{The definition of this free energy is unexpectedly subtle, due to the presence of the Arf fermionic SPT in (1+1)d, see e.g. \cite{Smith:2020rru}.}
\be
\cF_{A} =  \frac{N}{2}\log(2) \, , \ \ \ \cF_{V}(\Sigma = 0) = 0 \, .
\ee
The axial boundary condition can thus flow to the vector one, even at weak coupling. This is the known RG flow from the Andreev boundary condition to the standard boundary condition for a free Dirac fermion.

\begin{figure}[t]
    \includegraphics[width=1.00\linewidth]{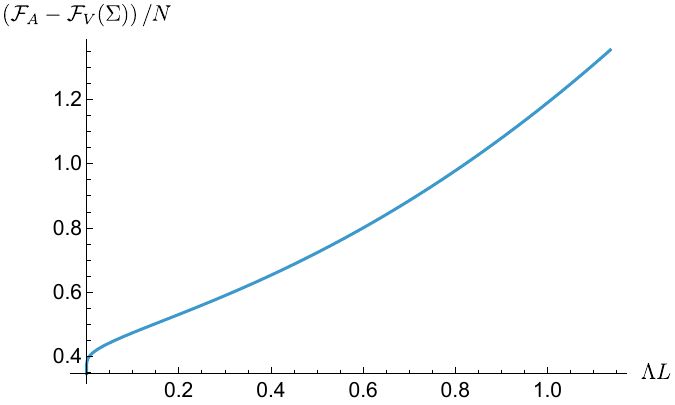}
    \caption{The large N free energy difference between massless (axial-preserving) and massive (vector-preserving) b.c. in the GN model.}
    \label{fig:GN}
\end{figure}

In \cite{Copetti:2023sya} it was predicted that the $|A,\eta\rangle$ b.c. is unstable at intermediate $\Lambda L$, and $|A,\eta \rangle$ has a discontinuous transition by merger-annihilation to the stable $|V,\Sigma\rangle$ boundary condition. As the free energy difference $\cF_A - \cF_V(\Sigma)$ remains positive (Figure \ref{fig:GN}), the claimed transition is allowed by the F-theorem.

\section{Conclusion and prospects}
We have proposed that the AdS free energy is monotonic under RG flows generated by boundary perturbations. This extends the scope of RG monotonicity theorems and suggests several promising directions for future work:

    \noindent \textbf{Longe Range CFTs:} Critical points of long-range statistical systems can often be realized as massive theories in AdS coupled to interacting boundary CFTs. In such cases, QFT in AdS---together with our AdS F-theorem---provides a powerful framework for identifying nontrivial long-range critical points and constraining RG flows between them, extending beyond the long-range Ising CFT discussed here.
    
    \noindent \textbf{Bulk RG:} Conventional defect monotonicity theorems break down when the bulk theory is non-conformal, since changing the RG scale simultaneously triggers bulk and boundary flows. Placing the theory in AdS offers a clean way to disentangle them. It would be interesting to investigate how bulk RG evolution (i.e.~varying $\Lambda L$) influences the AdS free energy. See \cite{Loparco:2024ibp, Abate:2024xyb, Abate:2024nyh} for related studies in de Sitter space. 

     \noindent \textbf{Confinement in AdS:} Recent work \cite{Aharony:2012jf,Copetti:2023sya,Ciccone:2024guw,Ciccone:2025dqx}  used AdS as a setting to study the onset of confinement, signaled by the instability of the weak-coupling Dirichlet boundary condition. At the transition point, the Dirichlet boundary condition  flows to a more stable boundary condition, which is conjectured to be the Neumann boundary condition. It would be interesting to use the AdS F-theorem to identify the endpoint of the associated boundary RG flow.
     
     \noindent \textbf{Flat Space Limit:} A further question is how the boundary RG flow and the AdS free energy behave in the flat-space limit. The flat-space limit of the boundary RG flow should control the IR divergence structure of flat-space amplitudes, but the precise connection remains to be clarified.

\vspace{2mm} \noindent \textbf{Acknowledgments.} We thank Chris Herzog, Edoardo Lauria, Marco Meineri, Ignacio Salazar Landea and Marco Serone for discussions. The work of C.C.~is supported by the STFC grant ST/X000761/1. The work of L.D.~is partially supported by INFN Iniziativa Specifica ST\&FI. C.C.~and L.D.~would like to thank the Isaac Newton Institute for Mathematical Sciences, Cambridge, for support and hospitality during the program ``Quantum field theory with boundaries, impurities, and defects", where work on this paper was undertaken. L.D.~and S.K.~thank the Yukawa Institute for Theoretical Physics and the organizers of the workshop ``Progress on Theoretical Bootstrap" for hospitality during the completion of this work.

 \setlength{\bibsep}{2pt plus 0.3ex}

 \bibliography{references.bib}

\newpage

\onecolumngrid

\begin{center}
    \LARGE \textbf{Supplemental Material}
\end{center}
\appendix
\section{Regularization of the free massive scalar}
In this Section we describe several methods to consistently regularize the massive free scale on both AdS and the sphere. In all cases it is convenient to study the derivative of the free energy with respect to the mass:
\be
\frac{\partial}{\partial m^2 L^2} F = \frac{1}{2 L^2} \int_{\cM} d^{d+1} x \sqrt{g} \langle \phi(x)^2 \rangle \, ,
\ee
with appropriate regularization of the composite operator $\phi^2$. For example, in AdS we can formally take \cite{Carmi:2018qzm}:
\be
\frac{\partial}{\partial m^2 L^2} F_{\AdS}(\pm) = \frac{\text{Vol}(\AdS)}{L^{d+1}} \frac{\Gamma(\frac{d}{2})}{8 \pi^{d/2+1}\Gamma(d)} \int_{\cC_\pm} d\nu \frac{1}{\nu^2 + d^2/4 +m^2L^2} \frac{\Gamma(d/2 + i \nu)\Gamma(d/2 - i \nu)}{\Gamma( i \nu)\Gamma(- i \nu)} \, .
\ee
while on the sphere:
\be
\frac{\partial}{\partial m^2 R^2} F_{S^{d+1}} = \frac{\text{Vol}(S^{d+1})}{2 R^{d+1}} \sum_{l=0}^\infty \frac{(d+2l)\Gamma(d+l)}{(4\pi)^{(d+1)/2} \Gamma(\frac{d+1}{2})\Gamma(l+1) } \frac{1}{l(l+d) + m^2 R^2} \, .
\ee
The IR divergence in AdS is proportional to the volume, and is regulated by substituting it with its regularized counterpart $\widetilde{\text{Vol}}(\AdS)$. The remaining sum/integrals are still UV divergent and need further regularization.

\vspace{2mm}\noindent\textbf{Pauli--Villars}
For $d=1,2$ Pauli--Villars (PV) is a feasible regularization method. Introducing a single PV field of mass $M_{\text{PV}}$ cures the UV divergences; for higher $d$ multiple PV fields are needed. For $d=1$ we find
\bea
\frac{\partial F_{\AdS}^{PV}(\pm)}{\partial (m^2 L^2)}
&= \frac{1}{4}\Big[
\psi\!\left(\frac{1}{2}-\sqrt{\frac{1}{4}+m^2 L^2}\right)
+ \psi\!\left(\frac{1}{2}+\sqrt{\frac{1}{4}+m^2 L^2}\right)
- \psi\!\left(\frac{1}{2}-\sqrt{\frac{1}{4}+M_{PV}^2 L^2}\right)
- \psi\!\left(\frac{1}{2}+\sqrt{\frac{1}{4}+M_{PV}^2 L^2}\right)
\\
&\qquad\qquad
\pm \pi \tan\!\left(\pi \sqrt{\frac{1}{4}+m^2 L^2}\right)
- \pi \tan\!\left(\pi \sqrt{\frac{1}{4}+M_{PV}^2 L^2}\right)
\Big] .
\eea

On the sphere,
\bea
\frac{\partial F_{S^2}^{PV}}{\partial (m^2 R^2)}
&= -\frac{1}{2}\Big[
\psi\!\left(\frac{1}{2}-\sqrt{\frac{1}{4}-m^2 R^2}\right)
+ \psi\!\left(\frac{1}{2}+\sqrt{\frac{1}{4}-m^2 R^2}\right)
\\
&\qquad\qquad
- \psi\!\left(\frac{1}{2}-\sqrt{\frac{1}{4}-M_{PV}^2 R^2}\right)
- \psi\!\left(\frac{1}{2}+\sqrt{\frac{1}{4}-M_{PV}^2 R^2}\right)
\Big] .
\eea
which leads to
\be
\frac{\partial \cF_{\AdS_2}(\pm)}{\partial (m^2 L^2)}
= \mp\,\frac{\pi}{4}\,\tan\!\left(\pi\sqrt{\tfrac{1}{4}+m^2 L^2}\right) .
\ee
For $d=2$, 
\be
\frac{\partial F_{\AdS}^{\text{PV}}(\pm)}{\partial (m^2 L^2)}
= \frac{1}{4}\!\left(\sqrt{1+M_{\text{PV}}^2 L^2}\;\mp\;\sqrt{1+m^2 L^2}\right) ,
\ee
and
\be
\frac{\partial F_{S^3}^{\text{PV}}}{\partial (m^2 R^2)}
= \frac{\pi}{4}\!\left[
\sqrt{1-M_{\text{PV}}^2 R^2}\;\cot\!\left(\pi\sqrt{1-M_{\text{PV}}^2 R^2}\right)
-\sqrt{1-m^2 R^2}\;\cot\!\left(\pi\sqrt{1-m^2 R^2}\right)
\right] ,
\ee
leading to
\be
\frac{\partial \cF_{\AdS_3}(\pm)}{\partial (m^2 L^2)}
= \mp\,\frac{1}{4}\,\sqrt{1+m^2 L^2}\, .
\ee

\vspace{2mm}\noindent\textbf{Dimensional Regularization}
Dimensional regularization also cures the UV divergences in a similar manner. We find
\be
\frac{\partial F_{\AdS}^{\text{dimreg}}(\pm)}{\partial (m^2 L^2)}
= \frac{\Gamma(-d)\,
\Gamma\!\left(\tfrac{d}{2}+\sqrt{\tfrac{d^2}{4}+m^2 L^2}\right)
\Gamma\!\left(\tfrac{d}{2}-\sqrt{\tfrac{d^2}{4}+m^2 L^2}\right)
\sin\!\left(\pi\left(\tfrac{d}{2}\mp\sqrt{\tfrac{d^2}{4}+m^2 L^2}\right)\right)}{2\pi}\,,
\ee
and
\be
\frac{\partial F_{S^{d+1}}^{\text{dimreg}}}{\partial (m^2 R^2)}
= -\,\frac{\Gamma(-d)\,\sin(\pi d/2)\,
\Gamma\!\left(\tfrac{d}{2}+\sqrt{\tfrac{d^2}{4}-m^2 R^2}\right)
\Gamma\!\left(\tfrac{d}{2}-\sqrt{\tfrac{d^2}{4}-m^2 R^2}\right)
\cos\!\left(\pi\sqrt{\tfrac{d^2}{4}-m^2 R^2}\right)}{\pi}\,.
\ee
To get a compact answer, it is useful to introduce the dimensionally continued free energy \cite{Giombi:2014xxa}:
\be
\cF_{\AdS}^{\text{dimreg}}
= - \sin\!\left(\frac{\pi d}{2}\right)\!\left(
F_{\AdS}^{\text{dimreg}}
- \frac{1}{2}\,F_{S^{d+1}}^{\text{dimreg}}\Big|_{R=iL}
\right).
\ee
This is equivalent to our $\cF_{\AdS}$, up to a rescaling by $2/\pi$ in even $d$. We find the finite answer
\be
\frac{\partial \cF_{\AdS}^{\text{dimreg}}(\pm)}{\partial (m^2 L^2)}
= \pm\,\frac{\Gamma(-d)\,\sin(\pi d)\,
\Gamma\!\left(\tfrac{d}{2}+\sqrt{\tfrac{d^2}{4}+m^2 L^2}\right)
\Gamma\!\left(\tfrac{d}{2}-\sqrt{\tfrac{d^2}{4}+m^2 L^2}\right)
\sin\!\left(\pi\sqrt{\tfrac{d^2}{4}+m^2 L^2}\right)}{4\pi}\,.
\ee
It can be checked explicitly that this matches the PV result in low dimensions.

\vspace{2mm} \noindent \textbf{Heat Kernel} The free energy computation on a sphere using the heat kernel was studied in \cite{Anninos:2020hfj}. Here we quote the result:
\begin{align}\label{eq:FSheat}
\begin{aligned}
F_{S^{d+1}}^\text{HK} =&- \int_{\epsilon}^{\infty} \frac{dt}{2\sqrt{t^2 - \epsilon^2}} \, \frac{1 + e^{-t}}{1 - e^{-t}} \, \frac{
e^{-\frac{d}{2}t + i\nu_{S} \sqrt{t^2 - \epsilon^2}} + e^{-\frac{d}{2}t - i\nu_{S} \sqrt{t^2 - \epsilon^2}}
}{
(1 - e^{-t})^d
},\\
\overset{\epsilon\to 0}{=}&- \int_{0}^{\infty} \frac{dt}{2t} \, \frac{1 + e^{-t}}{1 - e^{-t}} \, \frac{
e^{-\frac{d}{2}t + i\nu_{S} t} + e^{-\frac{d}{2}t - i\nu_{S} t}
}{
(1 - e^{-t})^d
}\,,
\end{aligned}
\end{align}
where $\nu_S\equiv\sqrt{m^2R^2-d^2/4}$ and $\epsilon$ is a smooth UV cut-off. After formally taking $\epsilon\to0$ and taking the $\nu_S$ derivative of the free energy
\begin{align}
\partial_{\nu_S}F_{S^{d+1}}^\text{HK}=& i\int_{0}^{\infty} \frac{dt}{2} \, \frac{1 + e^{-t}}{1 - e^{-t}} \, \frac{e^{-\frac{d}{2}t - i\nu_{S} t}-
e^{-\frac{d}{2}t + i\nu_{S} t}  
}{
(1 - e^{-t})^d
}\,,
\end{align}
it is possible to perform this integral by a change of integration variable to $x=e^{-t}$. We then get
\begin{align}\label{eq:derFSfinal}
\partial_{\nu_{S}}F_{S^{d+1}}^\text{HK}=\frac{\pi\nu_{S}}{\sin(\pi d) \Gamma(1+d)}\left[\frac{\Gamma(\frac{d}{2}+i\nu_{S})}{\Gamma(1-\frac{d}{2}+i\nu_{S})}+\frac{\Gamma(\frac{d}{2}-i\nu_{S})}{\Gamma(1-\frac{d}{2}-i\nu_{S})}\right]\,.
\end{align}
The UV divergence is cured when by analytic continuation in $d$.

We now generalize the result to AdS. We build on the results in \cite{Giombi:2013yva} and rewrite it using the $\epsilon$-regularization performed in \cite{Anninos:2020hfj}. The result is
\begin{align}
\begin{aligned}
F_{AdS_{d+1}}^\text{HK}&=-\int_{0}^{\infty} \frac{du}{2\sqrt{u^2+\epsilon^2}}\frac{1+e^{-u}}{1-e^{-u}}\frac{e^{-\nu\sqrt{u^2+\epsilon^2}-\frac{du}{2}}}{(1-e^{-u})^d}\\
&\overset{\epsilon=0}{=}-\int_{0}^{\infty} \frac{du}{2u}\frac{1+e^{-u}}{1-e^{-u}}\frac{e^{-\nu u-\frac{du}{2}}}{(1-e^{-u})^d}\,
\end{aligned}
\end{align}
where $\nu=\sqrt{m^2L^2+d^2/4}$. The integral can be made finite by analytic continuation in $d$ and can be computed by taking the $\nu$ derivative and by a change of variable $x=e^{-u}$:
\begin{align}
\partial_{\nu}F_{AdS_{d+1}}^\text{HK}=\int_{0}^{\infty} \frac{du}{2} \frac{1+e^{-u}}{1-e^{-u}}\frac{e^{-\nu u-\frac{du}{2}}}{(1-e^{-u})^d}=\frac{\nu\Gamma(-d)\Gamma(\frac{d}{2}+\nu)}{\Gamma(1-\frac{d}{2}+\nu)}\,.
\end{align}
The derivative of the regularized free energy of global AdS$_{d+1}$ can be computed using the heat kernel results, leading to Table \ref{tab:freescalar}.

\section{Regularization of the free massive fermion}
Free massive Dirac fermions in AdS$_{d+1}$ decay near the boundary with the following fall off (in Poincar\'e coordinates)
\begin{align}
\psi(z,x)\sim z^{\Delta_-}\psi_-(x)+z^{\Delta_+}\psi_+(x)\,,
\end{align}
where
\begin{align}
    \Delta_\pm=\frac d2 \pm |m|\,.
\end{align}
When
\begin{align}\label{eunitaritybound}
0\leq |m|\leq \frac 12
\end{align}
both modes are normalizable and can be chosen to be the dynamical ones. We use $\pm$ in different ways to denote the different boundary conditions where the fall offs are given by $z^{\Delta_\pm}$. The boundary condition $-$ is connected by the $+$ one by a bRG. We will use this fact to test the monotonicity of the free energy we define.

We employ point splitting to regulate the UV divergences of 
\begin{align}\label{epartialmZ}
   -  \int\! d^{d+1}x \sqrt{g}\  \langle \bar{\psi}\psi \rangle_\pm(x)=\partial_m  F(\pm)\,.
\end{align}
More specifically let us define the following embedding coordinates for AdS $X^A=(X^+,X^-,X^a)=\tfrac{1}{z}(1,x^2+z^2,x^a)$, with $X^\pm=X^0\pm X^{d+1}$, such that $X^2=-L^2$. We then define the chordal distance as
\begin{align}\label{ecordial}
    u=\frac{(X-Y)^2}{2L^2}\,.
\end{align}
With this we can write the propagator in AdS with $z^{\Delta_+}$ fall off \cite{Kawano:1999au}
\begin{align}
\langle\psi(z)\bar{\psi}(w)\rangle=-(\slashed{D}+\Gamma^0+m)\left[ \sqrt{\frac{z_0}{w_0}}\left(G_{\frac{d}{2}+m-\frac{1}{2}}(z,w)\mathcal{P}_-+G_{\frac{d}{2}+m+\frac{1}{2}}(z,w)\mathcal{P}_+\right)\right]\,,
\end{align}
where we use $\mathcal{P}_\pm=\tfrac{\mathbb{1} \pm\Gamma^0}{2}$, $\Gamma^0$ is defined in tangent space, and
\begin{align}
G_\Delta(z,w)=\frac{\Gamma\left(\frac{\Delta}{2}\right)\Gamma\left(\frac{\Delta+1}{2}\right)}{4\pi^{\frac{d+1}{2}}(u+1)^\Delta\Gamma\left(\Delta-\frac{d}{2}+1\right)} {}_{2}F_1\left(\frac{\Delta}{2},\frac{\Delta+1}{2},\Delta-\frac{d}{2}+1;\frac{1}{(1+u)^2}\right)\,.
\end{align}
The propagator with fall-off $z^{\Delta_-}$ can be obtained by an action of parity, by sending $m\rightarrow -m$ and by multiplying by $-1$ the final propagator.
With this we can compute the coincident limit of the propagator needed to obtain \eqref{epartialmZ}. We do this by setting $z_x=z_y$, which allows to have an expression as a function of $u$ only, and then take the $u\rightarrow 0$ limit.

\begin{figure}[htbp]
    \centering
    \includegraphics[width=0.7\linewidth]{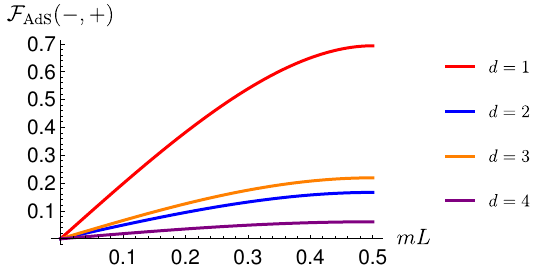}
\caption{Free energy difference for a massive Dirac fermion in AdS.
We focus on $m\geq 0$, within the unitarity bound of both boundary conditions \eqref{eunitaritybound}, and on $1\leq d\leq 4$.}\label{ffreefermion}
\end{figure}

We tested for $1\leq d\leq 8$ that both $\partial_m F(+)-\partial_m F(-)$ and $\partial_m \cF_{\text{AdS}_{d+1}}$ from equation \eqref{eq:Fads} are free of UV divergences.
We use propagators in dS$_{d+1}$ \cite{Schaub:2023scu} to obtain the one on $S^{d+1}$. We need to match the regulators for the divergences in AdS and on the sphere, since in $\cF_{\text{AdS}}$ these should cancel. We do so by employing the same cordial distance \eqref{ecordial}, but with the embedding space coordinates of the sphere, and the same type of coincident limit. We also take into account that, as we rotate $L_{\text{Sph}}\rightarrow e^{i\theta}L_{\text{AdS}}$, we have that for the sphere $u\rightarrow e^{-2i\theta} u$. We can then compute
\begin{align}
    F(-)-F(+)=\int_0^m dm'\ \left(\partial_{m'} F(-)-\partial_{m'} F(+)\right)
\end{align}
using the fact that the two boundary conditions coincide at $m=0$ and therefore $F(-)-F(+)\bigg|_{m=0}=0$. In Figure \ref{ffreefermion} we show $\cF_{\text{AdS}}(-,+)$, from \eqref{eq:Fads}, for $1\leq d\leq 4$ and the consistency with the expected behavior of this free energy under the bRG.

\section{Double Trace Deformation}
We review some salient aspects of the double-trace deformation for the free massive scalar in AdS. 

We consider $-d^2/4\leq m^2 L^2 < -d^2/4 + 1$, where both $\pm$ boundary conditions are available. A bulk scalar field $\phi(z,x)$ has the asymptotics:
\be
\phi(z_0,x) \simAdSzzero z_0^{\Delta_+} \varphi_+(x) + ... + z_0^{\Delta_-} \varphi_-(x) + ...
\ee
the ... here denote subleading contributions which are uniquely fixed in terms of $\varphi_-$ and $\varphi_+$ \cite{Skenderis:2002wp}.
Consider the $-$ b.c. $\varphi_+=0$. $\varphi_-\equiv \hat{\cO}_-$ is a boundary primary of dimension $\Delta_- = d/2 - \nu$, $\nu= \sqrt{d^2/4+m^2L^2} \in [0,1)$. The bulk stress tensor is given by:
\be
T^{\mu\nu} = \partial^\mu \phi \partial^\nu \phi - \frac{g^{\mu\nu}}{2}\left( (\partial \phi)^2 + m^2 \phi^2 \right) \, .
\ee
This object is however not the correct one that leads to charges that vanish in the boundary limit. This is due to the presence of a relevant scalar $\varphi_-^2$ in the boundary spectrum, which can be subtracted by considering the improvement:
\be
\delta T^{\mu\nu} = \frac{\beta}{2}\left( g^{\mu\nu} \Box - \nabla^\mu \nabla^\nu + R^{\mu\nu} \right) \phi^2 \, .
\ee
Alternatively, this corresponds to a preferred splitting between the bare mass and the conformal mass of the AdS scalar. From:
\bea
&\delta T^{\perp}_{~a} &&= -\frac{\beta}{2} (\partial_\perp + 1) \partial_a \phi^2 \, , \\
&\delta T^{\perp \perp} &&= -\frac{\beta}{2} \left[ d(\partial_\perp + 1) - z^2 \widehatBox\right] \phi^2 \, ,
\eea
it follows that requiring Weyl invariance fixes:
\be
\beta = \frac{\Delta_-}{2 \Delta_- + 1} \, .
\ee
For the massless scalar, this reproduces the conformal stress tensor in curved spacetime. The stress tensor bOPE now reads:
\be
T^{\perp a} \simAdSzzero \mathcal{O}(z_0^{d +1+ 2(1-\nu)}) \, , \qquad  T^{\perp \perp} \simAdSzzero - \frac{1}{2 (2 \Delta_- +1)} z_0^{d + 2(1-\nu)} \left[ \hat{\partial}_a \hat{\cO}_- \hat{\partial}^a \hat{\cO}_- - \frac{\Delta_-}{2 \Delta_- - d +2} \hat{\cO}_- \widehatBox \hat{\cO}_- \right] + ... \, .
\ee
which indeed is a primary for the $-$ bCFT \cite{Fitzpatrick:2011dm}. Notice that, as we exit the window in which the $-$ b.c. is stable the displacement becomes a relevant boundary deformation, as expected from general remarks on AdS boundary conditions \cite{Lauria:2023uca}. We now consider the boundary conditions giving rise to the double-trace deformation, namely:
\be
\varphi_+ = - (d - 2 \Delta_-)^{-1} f \, \hat{\cO}_- \, .
\ee
The leading asymptotics now are:
\bea
&T^{\perp }_{~ a} && \simAdS -z_0^{d} \frac{f}{2 \Delta_- + 1} \partial_a \left( \frac{\hat{\cO}_-^2}{2} \right) + ... \\
&T^{\perp \perp}  &&  \simAdS - z_0^{d} \frac{f}{2 \Delta_- + 1} \Delta_- (d - 2 \Delta_- - 1) \hat{\cO}_-^2 \, + ... \, .
\eea
In the notation of footnote \ref{foot:hattheta}, we thus have
\begin{equation}
\widehat{\Sigma}=~ -\frac{f}{2 \Delta_- + 1} \Delta_- (d - 2 \Delta_- - 1) \hat{\cO}_-^2  ,~~\widehat{\tau}^a_{~b}= -\frac{f}{2 \Delta_- + 1}   \frac{\hat{\cO}_-^2}{2} \delta^a_{~b}~.
\end{equation}
Combining these we find:
\be
\widehat{\Theta} = (d - 2 \Delta_-) f \frac{\hat{\cO}_-^2}{2}  \, .
\ee
where we recognize the classical beta function for the double trace coupling $f$.

\section{Flows between large $N$ bCFT.}
In recent work \cite{Copetti:2023sya}, the authors studied symmetry breaking and phase transitions of large $N$ models in two-dimensional AdS space. By relating the boundary RG flow with the bulk phase transition, the authors were able to find a quantitative boundary signal of bulk mass gap emerging. In this section, we study the $F_\text{AdS}$ theorem applied to these models and examine the interpretation of these phase transitions as boundary RG flows. 

\vspace{2mm} \noindent \textbf{The $O(N)$ model.}
First, we consider the $O(N)$ non-linear sigma model in AdS \cite{Carmi:2018qzm}. In a finite range of AdS radius $\frac{1}{4}<\Lambda L<1$, both the symmetry-preserving massive phase and the symmetry-breaking massless phase admit unitary conformal Dirichlet boundary conditions. Given the large $N$ effective action:
\be
\mathscr{S} = \int \frac{1}{2} (\partial \phi)^2 + \sigma\left( \phi^2 - \frac{N}{g^2} \right) + \frac{N}{2} \text{Tr} \log\left[ -\Box + 2 \sigma  \right] \, ,
\ee
and the gap equations therefrom:
\bea
\Sigma \, \Phi^i &= 0 \, , \qquad
\Phi^2 - \frac{1}{g^2} + \text{Vol}_{\text{AdS}_2}^{-1} \text{Tr} \left[ \frac{1}{- \Box + 2 \Sigma} \right] &= 0 \, ,
\eea
the two phases correspond to two solutions of the gap equations respectively:
\begin{equation}
    \begin{aligned}
        &\Sigma=0,\,\Phi^i\neq0,\quad \text{symmetry breaking, massless}\\
        &\Sigma\neq0,\,\Phi^i=0,\quad \text{symmetry preserving, massive}.
    \end{aligned}
\end{equation}
We use $F(s,\varphi)$ to denote the free energy of a phase with gap $s$ and vev $\varphi$. The free energy of the massless phase is simply
\begin{equation}
    F(0,\Phi^i)=F(0,0)+\int_0^{\Phi^i}d\varphi\,\frac{\partial}{\partial \varphi} F(0,\varphi)=F(0,0)
\end{equation}
because $\frac{\partial}{\partial \varphi} F(0,\varphi)=s\varphi=0$ for massless phases. The massive phase free energy is
\begin{equation}
    F(\Sigma,0)=F(0,0)+\int_0^\Sigma\,ds\,\frac{\partial}{\partial s} F(s,0)
\end{equation}
The derivative of the free energy with respect to the gap is
\begin{equation}
    \frac{\partial}{\partial s}F(s,0) = \text{Vol}_{\text{AdS}_2}N\left(- \frac{1}{g^2} + \,\text{tr} \left[ \frac{1}{- \Box + 2 s} \right]\right),
\end{equation}
where the dimensionally-regulated inverse propagator is
\be
\begin{aligned} \label{eq: invpropON}
&\text{Vol}_{\text{AdS}_2}^{-1}\text{Tr} \left[ \frac{1}{-\Box + 2 s} \right] = - \frac{1}{2\pi} \left[ \psi\left(\frac{1}{2} +\sqrt{\frac{1}{4} + 2 s}\right) + \gamma \right] \\ 
&+\frac{1}{2\pi}\left(- \frac{1}{\epsilon} + \gamma + \log(\mu L) + 2 \log(4\pi)\right) \, .
&
\end{aligned}
\ee
We absorb the divergence of the second line into the definition of the regularized coupling $1/g_\text{reg}^2\equiv1/g^2+1/(\pi\epsilon)-\gamma/\pi-2\log(4\pi)/\pi$ and $\Lambda\equiv\mu e^{-\pi/2g_\text{reg}^2}$, so the derivative becomes
\begin{equation}\label{eq:Fders}
    \frac{\partial}{\partial s}F(s,0) = \text{Vol}_{\text{AdS}_2}N\left(\frac{1}{2\pi}\log(\Lambda L)- \frac{1}{2\pi} \left[ \psi\left(\frac{1}{2} +\sqrt{\frac{1}{4} + 2 s}\right) + \gamma \right]\right).
\end{equation}
Using the regularized AdS${}_2$ volume $-2\pi$ and integrating \ref{eq:Fders} to get $F(\Sigma,0)$, we can plot $\left(F(0,\Phi^i)-F(\Sigma,0)\right)/N$ at any radius $\frac{1}{4}<\Lambda L<1$. See figure \ref{fig:ON}.

In the range $0<\Lambda L<\frac{1}{4}$, the symmetry preserving massive phase admits unitary conformal Neumann boundary condition while the symmetry breaking massless phase still admits the Dirichlet boundary condition as in $\frac{1}{4}<\Lambda L<1$. Now $\frac{\partial}{\partial s}F(s,0)$ receives additional contribution 
\begin{equation}
    \frac{\partial}{\partial s}F(s,0) = \text{Vol}_{\text{AdS}_2}N\left(\frac{1}{2\pi}\log(\Lambda L)- \frac{1}{2\pi} \left[ \psi\left(\frac{1}{2} +\sqrt{\frac{1}{4} + 2 s}\right) + \gamma \right]+\frac{1}{2}\tan\left(\pi \sqrt{\frac{1}{4} + 2 \Sigma} \right)\right).
\end{equation}
We can integrate it from the Breitenlohner-Freedman bound $\Sigma=-\frac{1}{8}$ to get the free energy of the Neumann massive phase. See the plots for $0<\Lambda L<\frac{1}{4}$ also in figure \ref{fig:ON}.

\begin{figure}[hbt]
    \centering
    \includegraphics[width=0.6\linewidth]{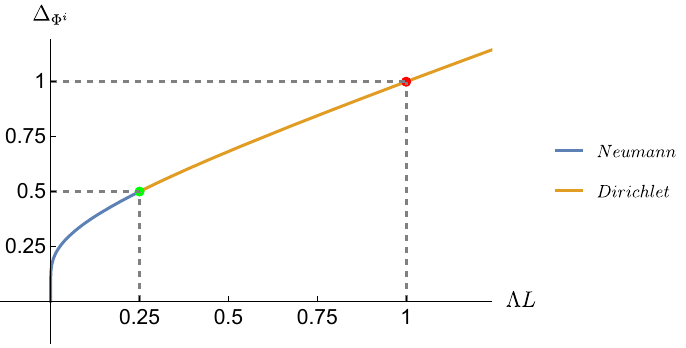}
    \caption{Conformal dimension of the boundary field of $\Phi^i$. This symmetry breaking operator remains relevant in the region $0<\Lambda L<1$.}
    \label{fig:Delta}
\end{figure}

From figure \ref{fig:ON}, in the range $0<\Lambda L<1$ where both the symmetry preserving and the symmetry breaking phase admit unitary conformal boundary conditions, the $F_\text{AdS}$ monotonicity constrains the RG flows between them and will only allow the symmetry preserving one to flow to the symmetry breaking one. This is consistent with the conformal dimension of the lightest fields that are charged under the boundary global symmetry. From figure \ref{fig:Delta} we can see that the boundary field of $\Phi^i$ is relevant in $0<\Lambda L<1$ and is irrelevant when $\Lambda L>1$. In $0<\Lambda L<1$, a deformation by $\Phi^i$ on the boundary will initiate the flow that breaks the $O(N)$ global symmetry and increases $F_\text{AdS}$. 

\begin{figure}[hbt]
    \centering
    \includegraphics[width=0.5\linewidth]{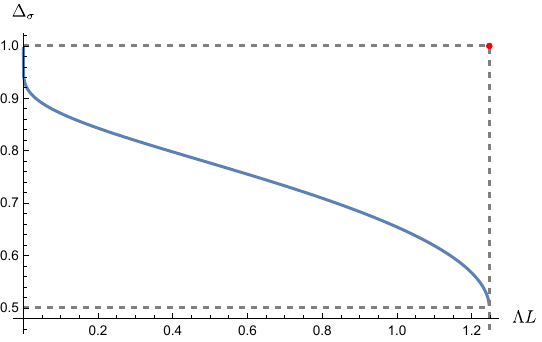}
    \caption{The conformal dimension of the axial $\mathbb{Z}_2$ odd operator $\sigma$ in the axial symmetric phase of the Gross-Neveu model. This operator is always relevant in the finite range of $\Lambda L$ where the axial symmetric boundary condition is believed to be stable. The red dot is where the axial neutral operator $\sigma^2$ becomes marginal and the boundary condition starts to be unstable.}
    \label{fig:Delta_sigma_new}
\end{figure}

\vspace{2mm} \noindent \textbf{The Gross-Neveu model.}
Another interesting model to look at is the Gross-Neveu model at large $N$. The large $N$ effective action is
\be
\mathscr{S} = \int \bar{\psi}_i \slashed{\nabla} \psi^i + \sigma \bar{\psi}_i \psi^i - \frac{N}{2g} \sigma^2 - N \, \text{Tr} \log \left[ \slashed{\nabla} + \sigma \right]
\ee
and the gap equation reads:
\be\label{eq: FermionGap}
\frac{\Sigma}{g} + \text{Vol}_{\text{AdS}_2}^{-1}\text{Tr} \left[ \frac{1}{\slashed{\nabla} + \Sigma} \right] = 0 \, .
\ee
We consider the vector symmetry preserving and the axial symmetry preserving boundary conditions respectively. Due to the mixed 't Hooft anomaly, there cannot be a boundary condition that preserves both symmetries. The two phases are:
\begin{equation}
    \begin{aligned}
        &\Sigma=0,\quad \text{axial symmetry preserving, massless}\\
        &\Sigma\neq0,\quad \text{vector symmetry preserving, massive}.
    \end{aligned}
\end{equation}
We use $F_V(s)$ to denote the free energy of the vector symmetry preserving phase with the plus boundary condition with gap $s$:
\begin{equation}
    F_V(\Sigma)=F_V(0)+\int_0^\Sigma\,ds\,\frac{\partial}{\partial s} F_V(s)
\end{equation}
The derivative of the free energy with respect to the gap is
\begin{equation}
    \frac{\partial}{\partial s}F_V(s) = \text{Vol}_{\text{AdS}_2}N\left(-\frac{s}{g} - \text{tr} \left[ \frac{1}{\slashed{\nabla} + s} \right]_+\right),
\end{equation}
where the dimensionally-regulated inverse propagator is
\begin{equation}
    \text{Vol}_{\text{AdS}_2}^{-1}\text{Tr} \left[ \frac{1}{\slashed{\nabla} +  \Sigma} \right]_+=-\frac{\Sigma }{\pi  \epsilon}+\frac{\Sigma  \left(-2\psi(|\Sigma|)
-\gamma +\log (4 \pi )\right)-\text{sgn}(\Sigma)}{2\pi }+O\left(\epsilon\right)
\end{equation}
We define $\Lambda=\mu e^{-\frac{\pi}{g_\text{reg}}}$ which absorbs the divergent term and some constants such that
\begin{equation}
    \frac{\partial}{\partial s}F_V(s) = \text{Vol}_{\text{AdS}_2}N\left(-\frac{s}{\pi}\log(\Lambda L)+\frac{2 s\,\psi(|s|)+\text{sgn}(s)}{2\pi}\right)
\end{equation}
In the free limit $\Lambda L\to 0$, both phases become the free fermions with appropriate conformal boundary conditions. Note that the full continuous axial symmetry can be restored at the free point. Conformal fermions in AdS can be mapped to conformal fermions on half space and thus their free energies are related to the defect $g$ function in flat space by $F=-\log g$. In particular, the vector symmetric defect has $\cF_{V}(\Sigma = 0) = 0$. The axial-symmetry free energy, however, does not change with $\Lambda L$ and always stays at the free fermion value $\cF_{A} =  \frac{N}{2}\log(2)$. The Gross-Neveu sphere free energy is canceled out in $\cF_V(\Sigma) - \cF_A$. In the end, we obtain figure \ref{fig:GN} in the main-text. The free energy of the vector symmetric phase is always larger than that of the axial symmetric phase, which is consistent with the possible flow away from axial symmetric phase induced by $\sigma$. See figure \ref{fig:Delta_sigma_new}.  

\end{document}